\documentclass[twocolumn,tighten]{aastex63}
\usepackage{lipsum}
\usepackage{amsmath}
\usepackage{float}

\usepackage{graphicx}
\usepackage{appendix}
\usepackage{tabularx,ragged2e}
\usepackage{booktabs}
\usepackage[rightcaption]{sidecap}

\begin{document}
\title{Bounds on velocity-dependent dark matter--proton scattering from Milky Way satellite abundance}

\shorttitle{DM--Proton Scattering Bounds from MW Satellites}
\shortauthors{Maamari et al.}

\correspondingauthor{Karime Maamari}
\email{maamari@usc.edu}

\author[0000-0002-7515-6208]{Karime Maamari}
\affiliation{Department of Physics and Astronomy, University of Southern California, Los Angeles, CA, 90089, USA}
\author[0000-0002-3589-8637]{Vera Gluscevic}
\affiliation{Department of Physics and Astronomy, University of Southern California, Los Angeles, CA, 90089, USA}
\author[0000-0003-1928-4667]{Kimberly K. Boddy}
\affiliation{Theory Group, Department of Physics, The University of Texas at Austin, Austin, TX 78712, USA}
\author[0000-0002-1182-3825]{Ethan O.~Nadler}
\affiliation{Kavli Institute for Particle Astrophysics and Cosmology and Department of Physics, Stanford University, Stanford, CA 94305, USA}
\author[0000-0003-2229-011X]{Risa H.~Wechsler}
\affiliation{Kavli Institute for Particle Astrophysics and Cosmology and Department of Physics, Stanford University, Stanford, CA 94305, USA}
\affiliation{SLAC National Accelerator Laboratory, Menlo Park, CA 94025, USA}

%%%%%%%%%%%%%%%%%%%%%%%%%%%%%%%%%%%%%%%%%%%%%%%%%%%%%%%%%%%%%%%%%%%%%%%%%%%%%%%%%%%%%%%%%%%%%%%%%%%%%%%%%%%%%%%%%%%%%%%%%%%%%%%%%%%%%%%%

\begin{abstract}
We use the latest measurements of the Milky Way satellite population from the Dark Energy Survey and Pan-STARRS1 to infer the most stringent astrophysical bound to date on velocity-dependent interactions between dark matter particles and protons.
We model the momentum-transfer cross section as a power law of the relative particle velocity $v$ with a free normalizing amplitude, $\sigma_\text{MT}=\sigma_0 v^n$, to broadly capture the interactions arising within the non-relativistic effective theory of dark matter-proton scattering.
The scattering leads to a momentum and heat transfer between the baryon and dark matter fluids in the early Universe, ultimately erasing structure on small physical scales and reducing the abundance of low-mass halos that host dwarf galaxies today.
From the consistency of observations with the cold collisionless dark matter paradigm, using a new method that relies on the most robust predictions of the linear perturbation theory, we infer an upper limit on $\sigma_0$ of $1.4\times10^{-23}$, $2.1\times10^{-19}$, and $1.0\times10^{-12}\ \mathrm{cm}^2$, for interaction models with $n=2,4,6$, respectively, for a dark matter particle mass of $10\ \mathrm{MeV}$. 
These results improve observational limits on dark matter--proton scattering by orders of magnitude and thus provide an important guide for viable sub-GeV dark matter candidates.
\end{abstract}

\reportnum{UTTG-13-2020}

%%%%%%%%%%%%%%%%%%%%%%%%%%%%%%%%%%%%%%%


\keywords{\href{http://astrothesaurus.org/uat/353}{Dark matter (353)}; \href{http://astrothesaurus.org/uat/1049}{Milky Way dark matter halo (1049)}; \href{http://astrothesaurus.org/uat/574}{Galaxy abundances (574)}}

%%%%%%%%%%%%%%%%%%%%%%%%%%%%%%%%%%

\section{Introduction} 
\label{sec:introduction}

After decades of versatile experimental searches, observations of the Universe remain the sole source of evidence for dark matter (DM).
Identifying its nature amounts to understanding a major constituent of matter and thus inspires investigations across different fields of physical science.
As the laboratory bounds on the most popular theoretical candidate models grow stronger, cosmological and astrophysical observations have emerged as an alternative and complementary probe of DM microphysics (for reviews, see \citealt{BuckleyPeter,Drlica-Wagner190201055,GluscevicAliHaimoud, Grin}).

The standard model of cosmology assumes cold dark matter (CDM) whose non-gravitational interactions are observationally insignificant.
Nearly all deviations from pure CDM considered in the current literature affect the way matter is distributed in the Universe, including warm DM (WDM; \citealt{Schneider:2016uqi,Abazajian:2017,Adhikari:2017}), fuzzy DM (FDM; \citealt{Hu:2000,Hui:2017}), self-interacting DM (SIDM; \citealt{Tulin:2018}), DM interacting with dark radiation (\citealt{CyrRacine:2015ihg}), and with Standard Model particles (IDM; \citealt{Escudero:2018,DvorkinBlum,GluscevicBoddy, BoddyGluscevic,NadlerGluscevic,NadlerDES}).
One of the original incentives to consider beyond-CDM models was the perceived ``missing satellites problem''---the apparent mismatch between the observed satellite galaxies orbiting the Milky Way and their predicted population from CDM simulations of structure formation \citep{Klypin, Moore}. 
Different properties of DM were invoked to account for the apparent mismatch, including appreciable free streaming (in WDM models), a macroscopic de Broglie wavelength (in FDM models), and particle interactions (in SIDM and IDM models).
Virtually all of them suppress the formation of low-mass DM halos and reduce the abundance of galaxies that would inhabit them.
However, a more recent census of faint galaxies in our galactic neighborhood, combined with advanced modeling of the galaxy--halo connection, has shown consistency between the CDM predictions and the observed satellite abundance, down to a halo mass of $\sim 10^{8}\ M_{\rm{\odot}}$  \citep{Kim:2018,Jethwa:2018,Newton:2018,Nadler:2019,Nadler:2020}. 
With this new development, measurements of the Milky Way satellite population can be reinterpreted to place stringent bounds on the microphysics of DM.

In this study, we focus on a scenario in which DM elastically scatters with normal matter (baryons), altering matter perturbations  in the early Universe and consequently reducing the present-day population of small galaxies.
We further rely on the concordance between the CDM predictions and measurements of the Milky Way satellite abundance from the Dark Energy Survey (DES) and Pan-STARRS1 (PS1; \citealt{Drlica-Wagner191203302,Nadler:2020,NadlerDES}) to place the most stringent astrophysical bounds on a variety of velocity-dependent interactions between DM particles and protons.

In a previous pilot study \citep{NadlerGluscevic}, we developed a method to constrain velocity-independent scattering only, which we later applied to DES data \citep{NadlerDES}. Here, we generalize our analysis to include a whole class of velocity-dependent interaction models; this generalization has required a new approach to quantifying the impact of DM interactions on the satellite population.
This method relies on predicting the most robust features of the matter transfer function in IDM cosmology and relating those features to the present-day abundance of low-mass halos. 
It does not necessitate precise modeling of the intricacies of galaxy formation within IDM.
As such, it is only suited for placement of conservative upper bounds on the interaction cross section. 
Even so, the upper bounds we obtain are 3--5 orders of magnitude more stringent than the previous observational limits.
They are also the first near-field cosmological limit on velocity-dependent DM--baryon interactions.

We address the same low-energy physics---and the same DM parameter space---as direct detection experiments.
However, as with most other observational approaches, it is particularly well-suited for probing relatively large interaction cross sections and sub-GeV particle masses, outside the target sensitivity of most nuclear-recoil-based underground experiments (e.~g.~\citealt{supercdms2019,xenonnt2020,EmkenKouvaris}).
It is thus directly complementary to laboratory searches for DM interactions with Standard Model particles and substantially reduces the allowed parameter space for IDM models.

The paper is organized as follows. 
In Section \ref{sec:theory}, we review the theoretical models of DM--proton scattering and their effects on observations.
In Section \ref{sec:observations}, we review the observational constraints on the Milky Way satellite galaxy population.
In Section \ref{sec:method}, we describe our approach to inferring upper limits on the interaction cross section from the measured abundance of the Milky Way satellites.
In Section \ref{sec:results}, we present our results.
We discuss our findings and conclude in Section \ref{sec:conclusion}.
Throughout, we adopt the following cosmological parameters: DM density~$\Omega_\text{dm} h^2 = 0.1153$, baryon density $\Omega_\text{b} h^2 = 0.02223$, radiation density $\Omega_\text{rad}\approx9.23\times10^{-5}$, the Hubble constant $h=0.6932$, optical depth to reionization $\tau_\text{reio} = 0.081$, the amplitude of the scalar perturbations $A_s = 2.464\times10^{-9}$, and the scalar spectral index $n_s = 0.9608$; we set $c=k_B=1$.\footnote{The parameter values are chosen to be consistent with those used in \citealt{NadlerDES}.}  
%%%%%%%%%%%%%%%%%%%%%%%%%%%%%%%%%%%%%%%%%%%%%%

\section{Theory}
\label{sec:theory}

We consider elastic scattering between DM particles and protons that predominantly takes place in the early Universe.\footnote{For simplicity, we ignore scattering with helium. This ensures that our bounds are conservative, as the inclusion of helium may only slightly improve them \citep{BoddyGluscevic}.} 
We consider any scattering process with a
momentum-transfer cross section of the form $\sigma_\text{MT} = \sigma_0 v^n$, where $v$ is the relative particle velocity and $\sigma_0$ is a free parameter of the model; we focus on power-law index values $n\in\{0, 2, 4, 6\}$ and consider a range of DM particle masses $m_\chi\in\left[15 \text{ keV},100 \text{ GeV}\right]$.\footnote{For $m_\chi$ much greater than a proton mass, the constraints we derive scale as $\sigma_0/m_\chi$ \citep{BoddyGluscevic}; for thermally-produced DM with masses below $\sim$10 keV, bounds on WDM apply \citep{Irsic:2017ixq}; other cosmological limits may apply in specific cases at masses $\lesssim$MeV, as discussed in Section \ref{sec:conclusion}.} 
We choose this empirical parameterization and values of $n$ because they are representative of wide variety of relativistic DM models which can be described by a low-energy effective field theory of DM scattering with nucleons, broadly considered in DM searches~\citep{Anand:2013yka,Fitzpatrick:2010br}.
The models are represented here by an appropriate choice of $n$ \citep{BoddyGluscevic}.
For example, $n=0$ represents a cross section with no velocity dependence and corresponds to a spin-independent or spin-dependent contact interaction, well-studied in context of direct detection; $n=2$ arises at leading order from DM with an electric dipole moment, induced by a heavy mediator that kinetically mixes with the photon~\citep{Fitzpatrick1302004}.
Aside from its connection to particle theory, the power-law parameterization is sufficient to fully capture the effects of scattering on structure formation and thermal history of the Universe, and is thus adopted as a standard approach in observational searches for DM interactions (e.~g.~\citealt{DvorkinBlum,GluscevicBoddy,BoddyGluscevic,Xu189710,SlatyerWu,BoddyGluscevicPoulin}).

\begin{figure}
\centering
    \hspace*{-5mm}
    \includegraphics[width=0.95\linewidth]{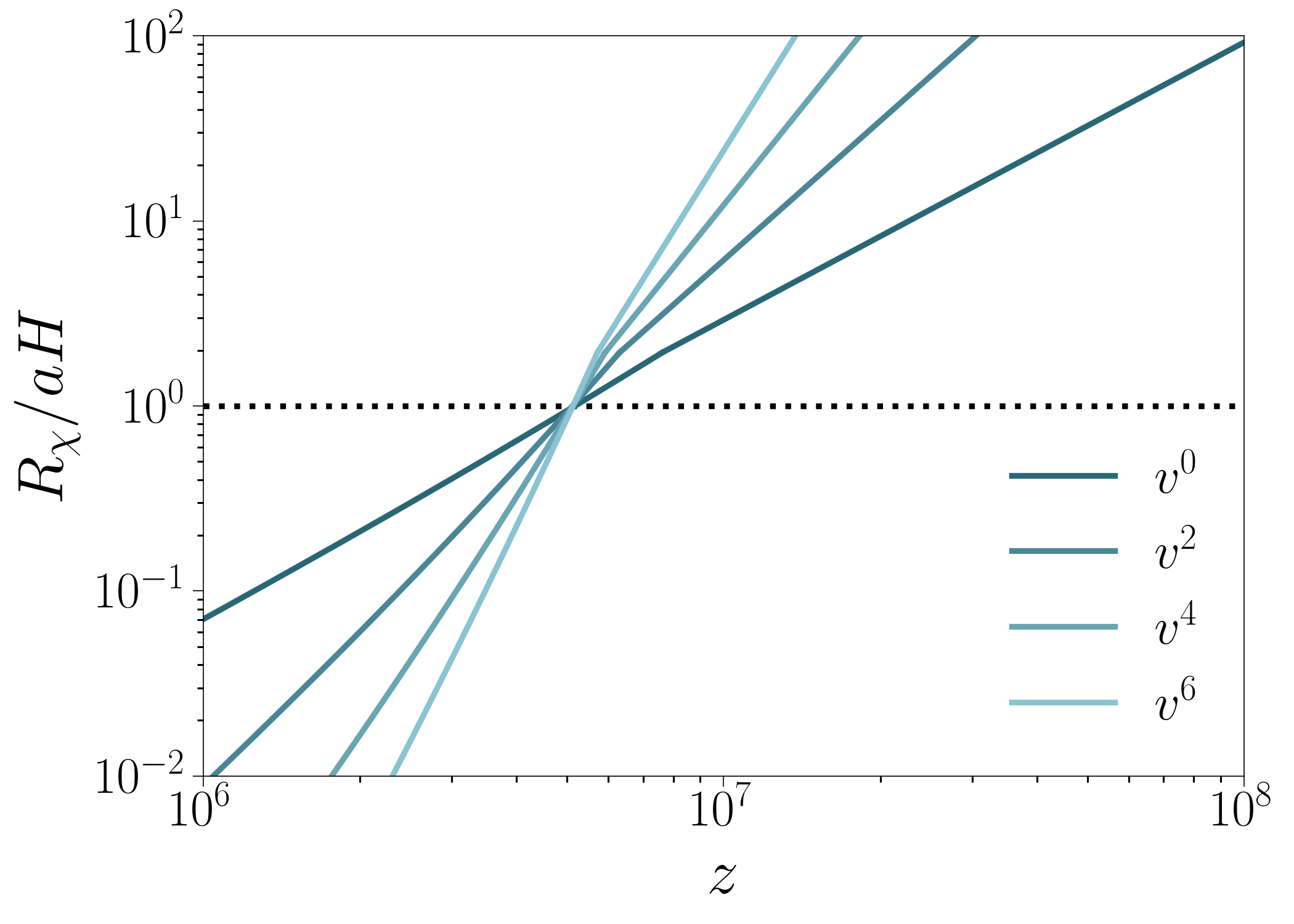}
    \caption{Redshift evolution of the momentum-transfer rate $R_\chi$ between DM and protons, normalized to the Hubble rate $aH$, for IDM models with a power-law dependence of the momentum-transfer cross section on relative particle velocity $v$. The DM mass is set to 1 MeV, and cross sections are normalized to the analytic estimates from Table \ref{tab:limits}. For all models considered, IDM scattering is only significant at high redshift.
    }
    \label{fig:Rx}
\end{figure}

\begin{figure*}
\centering
    \hspace*{-5mm}
    \includegraphics[width=0.46\textwidth]{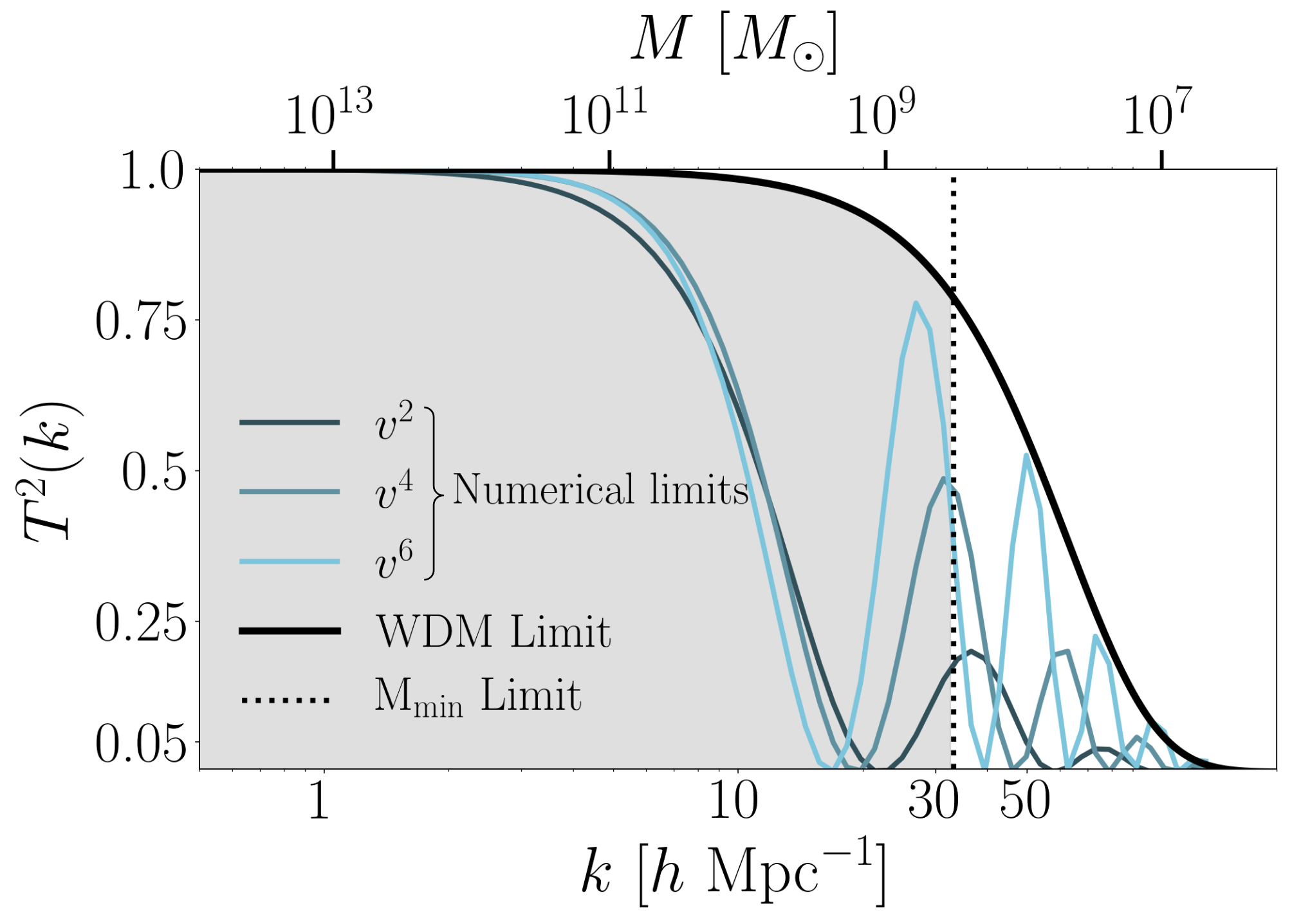}
    \includegraphics[width=0.46\textwidth]{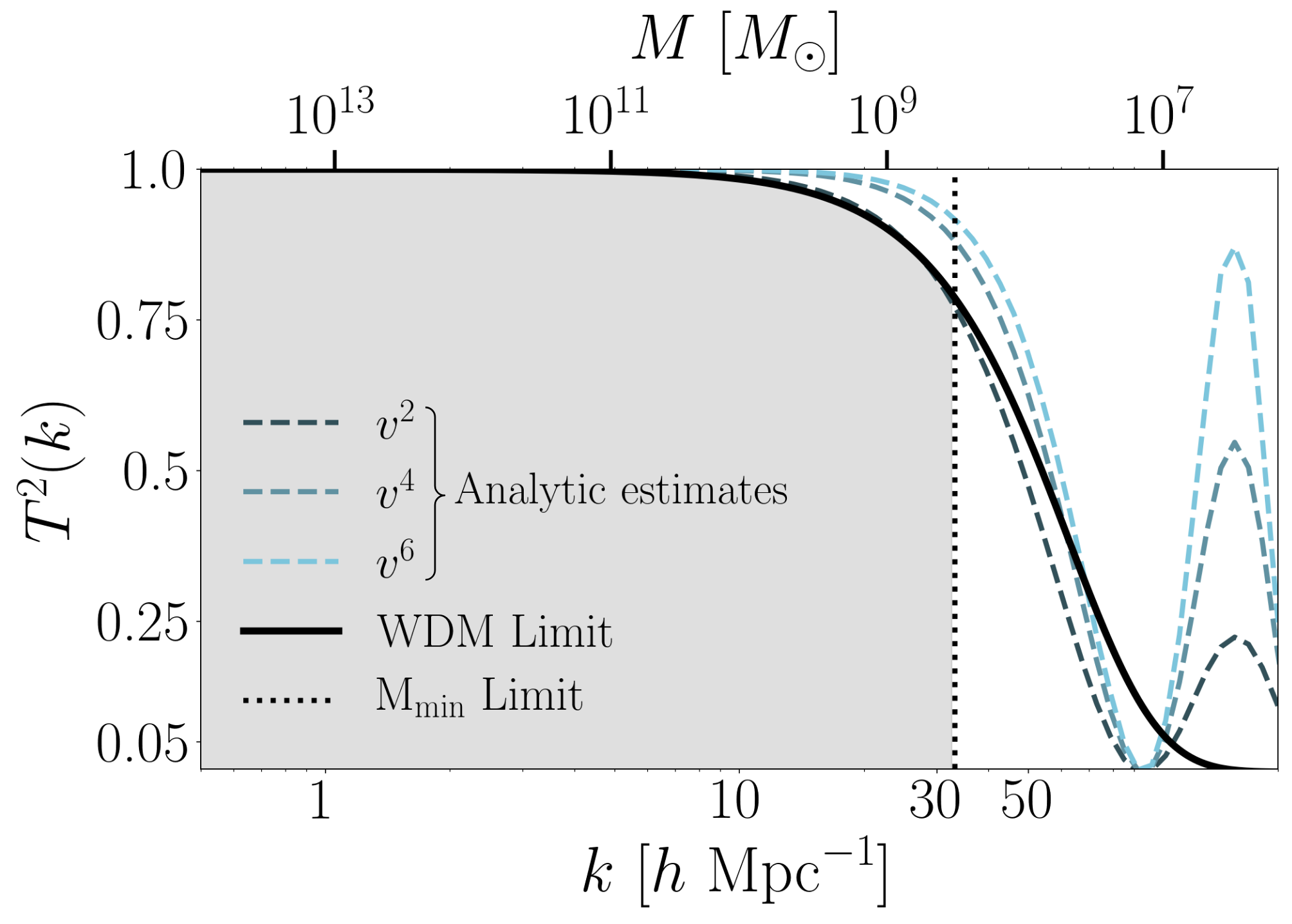}
    \caption{Square of the matter transfer function $T^2(k)$---the ratio of the linear matter power spectrum for a beyond-CDM cosmology to that of the standard CDM cosmology.
    The solid black line corresponds to WDM at the current lower mass limit of $6.5$ keV from the DES and PS1 satellite-abundance measurements \citep{NadlerDES}. The vertical dotted line corresponds to the upper limit on the minimum mass of DM halos inferred to host observed satellite galaxies, approximately mapped onto $k$-space, from the same study. Colored lines correspond to velocity-dependent DM-proton scattering for various power-law velocity dependencies of the momentum-transfer cross section, for a DM mass of 1 MeV.
    \textit{Left:} The coefficients of the IDM momentum-transfer cross sections are set to their upper bounds determined using our numerical approach in Section \ref{sec:method}. Larger cross sections produce a more prominent suppression, pushing the power cutoff to lower values of $k$, and are inconsistent with the data. \textit{Right:} The IDM cross sections are set to the values determined by the analytic estimates in Section \ref{sec:method}. Smaller cross sections are consistent with the data at most dark matter masses.} 
    \label{fig:cutoff}
\end{figure*}

In an IDM cosmology, DM--baryon scattering leads to heat and momentum transfer between the cosmological fluids, smoothing out small-scale density perturbations through collisional damping.  
The momentum-transfer rate $R_\chi$ and the heat-transfer rate $R'_\chi$ are proportional to $\sigma_0$, and their redshift evolution is largely dictated by the evolution of the relative particle velocities \citep{DvorkinBlum, GluscevicBoddy}.
Since particle velocities are primarily sourced by thermal motions in the early Universe ($z\gtrsim 10^4$), the associated $R_\chi$ evolves monotonically with redshift $z$, as the Universe cools (Figure~\ref{fig:Rx}). 
For models with $n\ge 0$, DM decouples from protons well before cosmic recombination and deep into this regime \citep{Dvorkin890212,BoddyGluscevic}.
In such cases, the interactions affect structure today primarily by means of suppressing the linear matter power spectrum $P(k)$ at small scales (large wave numbers $k$), early on in cosmic history; the square of the transfer function $T^2(k)\equiv P(k)/P_\text{CDM}(k)$ (the ratio between the IDM and the CDM power spectrum) features a cutoff, shown in Figure~\ref{fig:cutoff} in colored lines.
Models with $n<0$, on the other hand, feature scattering at late times, after structure formation commences.
Their effects are more challenging to compute and we leave their consideration for a future study. 
%%%%% Rx %%%%%
%%%%%%%%%%%%%%%%%%%%%%%%%%%

For $n\ge0$, DM scattering affects physical scales that enter the particle horizon prior to DM--baryon decoupling.
For the $n=0$ case of a velocity-independent interaction, the resulting cutoff in $T^2(k)$ is the main signature of IDM physics.
However, for the velocity-dependent interactions with $n>0$, there are also prominent ``dark acoustic oscillations'' (DAO; see \citealt{CyrRacine:2015ihg}) that appear at scales below the cutoff, due to the tight coupling between the photon--baryon fluid and the DM fluid at early times (Figure \ref{fig:cutoff}).
In both cases, the IDM-induced suppression of small-scale density perturbations ultimately leads to a decrement in the abundance of low-mass halos that host dwarf galaxies, as compared to the CDM cosmology. 

To forward-model a population of galaxies in a beyond-CDM cosmology and confront it with observations in principle requires a suite of fully consistent cosmological simulations, including beyond-CDM physics.
Such simulations are computationally expensive and were only performed for certain sets of beyond-CDM scenarios that feature a suppression of $P(k)$, most notably WDM, SIDM, and ETHOS models \citep{Schneider11120330,Angulo13042406,Lovell13081399,Bose160407409,CyrRacine:2015ihg,Murgia170407838}.
In \cite{NadlerGluscevic}, we utilized the fact that the velocity-independent DM-proton scattering (with $n=0$) suppresses $P(k)$ in a way that resembles WDM, and used the mapping between the parameters of the two models to derive bounds on this specific IDM case.
However, the mapping between WDM and IDM breaks down for velocity-dependent ($n>0$) scattering we focus on here because of the large DAO features.
In fact, the IDM transfer function does not straightforwardly map onto any other previously explored scenario with a power cutoff.
For this reason, we develop and apply a new method that relates the bounds on the matter transfer function, inferred from DES and PS1 measurements, to the most robust features of $T^2(k)$ within IDM with velocity-dependent scattering.
%%%%% Cutoffs %%%%%
%%%%%%%%%%%%%%%%%%%%%%%%%%%
%%%%%%%%%%%%%%%%%%%%%%%%%%%%
\section{Observations}
\label{sec:observations}

We use the recent measurements of the Milky Way satellite population by DES and PS1 \citep{Drlica-Wagner191203302}, and their inferred bounds on the matter transfer function $T(k)$.
The bounds are based on a probabilistic inference that combines i) models for satellite detectability in the relevant survey footprints \citep{Drlica-Wagner191203302}, ii) high-resolution DM--only simulations \citep{Mao150302637} chosen to match the observed characteristics of the Milky Way system, and iii) an empirical model of the galaxy--halo connection \citep{Nadler:2019,Nadler:2020}. 
By performing mock observations of the satellite populations and statistically comparing them to the luminosity, size, and radial distributions of the observed satellites, the model---including suppression of the subhalo mass function---is fit to the data using a Markov Chain Monte Carlo approach \citep{Nadler:2020,NadlerDES}.

The results are cast in terms of the bounds on models that suppress $T^2(k)$, notably a lower limit on the thermal-relic WDM mass of 6.5 keV at $95\%$ confidence \citep{NadlerDES}. 
The corresponding WDM $T^2(k)$ is shown as the solid black line in Figure~\ref{fig:cutoff}.
The same results were also cast as an upper limit on the minimum mass of halos that host observed satellite galaxies, $M_\text{min}<3.2\times 10^8$ $M_\odot$.
The corresponding comoving wave number $k_\text{crit} = 33.2\ h\ \mathrm{Mpc}^{-1}$, given by
\begin{equation}
    M_\text{min}= \frac{4\pi}{3}\Omega_\text{dm} \bar\rho \left(\frac{\pi}{k_\text{crit}}\right)^3,
    \label{eq:Mmin}
\end{equation}
represents the largest $k$ effectively probed by these data, shown as the vertical dotted line in Figure \ref{fig:cutoff}. $\bar\rho=4.75\times10^{-6}$ is the mean density of the Universe today and $\lambda_\text{crit}=2\pi/k_\text{crit}$ is the comoving size of perturbations giving rise to halos of average mass $M_\text{min}$.

%%%%%%%%%%%%%%%%%%%%%%%%%%%%%
\vspace{0.2cm}
\section{Method} 
\label{sec:method}

The WDM transfer function corresponding to the WDM mass limit, together with the minimum-halo-mass limit, in Figure \ref{fig:cutoff}, delineates the allowed region for $T^2(k)$: a viable beyond-CDM model must not suppress $T^2(k)$ more than the WDM model in this Figure, unless the suppression occurs beyond $k_\text{crit}$, where the data has no constraining power.
The implications to IDM are as follows: 
\begin{enumerate}
    \item Since $T^2(k)$ at the thermal-relic-WDM mass limit delineates the maximum suppression tolerated by current data, IDM models that produce a more suppressed $T^2(k)$ are \textit{inconsistent} with the data.
    
    \item $M_\text{min}$ is the minimum mass of halos whose abundance is demonstrably consistent with CDM; IDM models that exclusively alter the abundance of \textit{lower-mass} halos are \textit{consistent} with the data.
\end{enumerate}
We incorporate these points within a numerical approach and an analytic estimate, described below, to translate the $M_\text{min}$ and WDM mass bounds described in the previous Section into a bound on IDM. 
Before proceeding, we highlight an important distinction between them: the numerical approach yields an upper bound on $\sigma_0$, while the analytic estimate does \textit{not} provide upper bounds by itself; rather, it roughly \textit{estimates} a potential maximal improvement over the numerical limit, if full forward modeling of the satellite population is applied to the same data. 

%%%%%%%
\subsection{Numerical limits}

For each $n$, we compute the range of $\sigma_0$ for which $T^2(k)$ is strictly \emph{more} suppressed than the ruled-out thermal-relic-WDM model, as illustrated in the left panel of Figure~\ref{fig:cutoff}. 
To compute $T^2(k)$ for a given $\sigma_0$, $m_\chi$, and $n$, we use the modified Boltzmann code CLASS \citep{class} developed for IDM cosmology in \cite{BoddyGluscevic}.
We identify the value of $\sigma_0$ for which the transfer function, including its DAO features, lies entirely below the WDM limit.\footnote{We ensure that this condition holds down to a sufficiently small scale, $k<130\text{ }h\, \text{Mpc}^{-1}$.}
Finally, we repeat the procedure for each $m_\chi$ and $n$, obtaining $\sigma_0(m_\chi|n)$ as our numerical upper limit. 

This procedure produces robust upper bounds, because all DAO features of IDM lie strictly below the limit on WDM transfer function at each individual $k$ value, for all higher values of $\sigma_0$.
In other words, larger values of $\sigma_0$ produce a more prominent suppression in $T^2(k)$ and are thus excluded by data (at $>95\%$ confidence).\footnote{Our numerical limit is at $>95\%$ confidence, since the thermal-relic WDM limit it corresponds to is at $95\%$ confidence. However, since we did not perform likelihood analysis, we refrain from quantifying the confidence level exactly. This method is similar in spirit to the sterile neutrino analysis of \cite{Schneider:2016uqi}.}
In reality, the decrement of power in IDM at the numerical upper limit of $\sigma_0$ is already so prominent (left panel of Figure~\ref{fig:cutoff}), that the data is likely even more constraining. 
However, non-linear evolution of structure at such small scales makes it difficult to improve the limit on the basis of linear-theory considerations, without running IDM cosmological simulations. 
Nonetheless, our numerical limit presents a tremendous improvement over the other observational bounds on IDM, as quantified in Section \ref{sec:results}.

%%%%%%%%
%%%%%%%%%%%%%%%%%
\subsection{Analytic estimates}

As noted before, the analytic--limit prescription of \cite{NadlerGluscevic} does not strictly apply in generic IDM models.
We consider it here only as a rough \textit{estimate} for how much our numerical limits could be improved in principle.
We start by noting that IDM scattering affects matter perturbations until DM and baryons decouple, at $z_\text{dec}$.
Following \cite{NadlerGluscevic}, we find $z_\text{dec}$ by setting
\begin{equation}
    aH = R_\chi\big\rvert_{z_\text{dec}},
    \label{eq:decoupling}
\end{equation}
where \citep{BoddyGluscevicPoulin}
\begin{equation}
\small
    R_{\chi} = \frac{N_n a\rho_bY_p\sigma_0}{m_{\chi}+m_p}\Big(\frac{T_b}{m_p}+\frac{T_{\chi}}{m_{\chi}}\Big)^{\frac{1+n}{2}},
    \label{eq:Rx}
\end{equation}
\normalsize
$N_n=2^{5+n/2}\Gamma(3+n/2)/(3\sqrt{\pi})$, $a$ is the scale factor, $Y_p=0.75$ is the proton mass fraction, $\rho_b$ is the baryon energy density, and $m_p$ is the proton mass.
During radiation domination, the temperature of baryons evolves as $T_b=T_0(1+z)$, where $T_0=2.73$ K. 
The temperature of the DM fluid $T_{\chi}$ is strongly coupled to $T_b$ until thermal decoupling at $z_\text{th}$, and afterwards evolves adiabatically, $T_\chi=T_0(1+z)^2/(1+z_\text{th})$.
Thermal decoupling occurs when the heat transfer rate, $R_{\chi}^{\prime} \equiv R_{\chi}m_{\chi}/(m_{\chi}+m_p)$, matches the Hubble rate, $aH = R_{\chi}^{\prime}\big\rvert_{z_\text{th}}$.
Substituting Eq.~\eqref{eq:Rx} into the Eq.~\eqref{eq:decoupling}, we get $z_\text{dec}(\sigma_0| m_\chi,n)$.

We can further estimate a critical comoving scale below which collisional damping suppresses the matter transfer function; this scale corresponds to the size of the particle horizon at $z_\text{dec}$, given by
\begin{equation}
    k_\text{crit} \equiv 2aH\big\rvert_{z_\text{dec}} \approx 2 H_0z_\text{dec}\sqrt{\Omega_\text{rad}}.
    \label{eq:kcrit}
\end{equation} 
Substituting $z_\text{dec}(\sigma_0| m_\chi,n)$ in Eq.~\eqref{eq:kcrit}, we obtain $k_\text{crit}(\sigma_0| m_\chi,n)$.
Finally, we use Eq.~\eqref{eq:Mmin}, to relate $\sigma_0$ to the mean mass of the smallest halos affected by IDM physics, for a given $n$ and $m_\chi$. 
We note that a particular value of $k_\text{crit}$ translates to a different amount of suppression in $T^2(k)$, depending on the interaction model; however, this analytic prescription does not predict the amount of suppression.  
We also note that $M_\text{min}=3.2\times 10^8$ $M_\odot$ roughly corresponds to $z_\text{dec}\approx 4\times 10^7$ (the intersection point in Figure \ref{fig:Rx}).

The benefit of the analytic calculation for $n>0$ is that it provides a rough estimate of the largest mass at which halo abundances are affected by IDM, for a given $\sigma_0$.
In other words, the values of $\sigma_0$ that satisfy $M_\text{crit}(\sigma_0| m_\chi,n)<M_\text{min}$ only affect halos of masses below the current detection threshold. 
As such, they are largely allowed by the current data.
For illustration, the right panel of Figure~\ref{fig:cutoff} shows transfer functions for all our IDM models, where $\sigma_0$ is set using the analytic estimate.
The corresponding $T^2(k)$ curves present outer envelopes of the ``disallowed'' (shaded) region for most values of $k$.
We thus expect that the analytically--estimated bounds roughly capture maximal improvement that can be obtained with detailed forward modeling of the same data; however, this is a rough estimate that only holds true for some DM masses, as we show in the following.
%%%%%%%%%%%%%%%%%%%%%%%%%%%%%%%%%%%%%%%%%%

\section{Results}
\label{sec:results}

%%%%%%%%% LIMITS %%%%%%%%% 
\begin{figure*}
    \centering
    \includegraphics[width=0.475\textwidth]{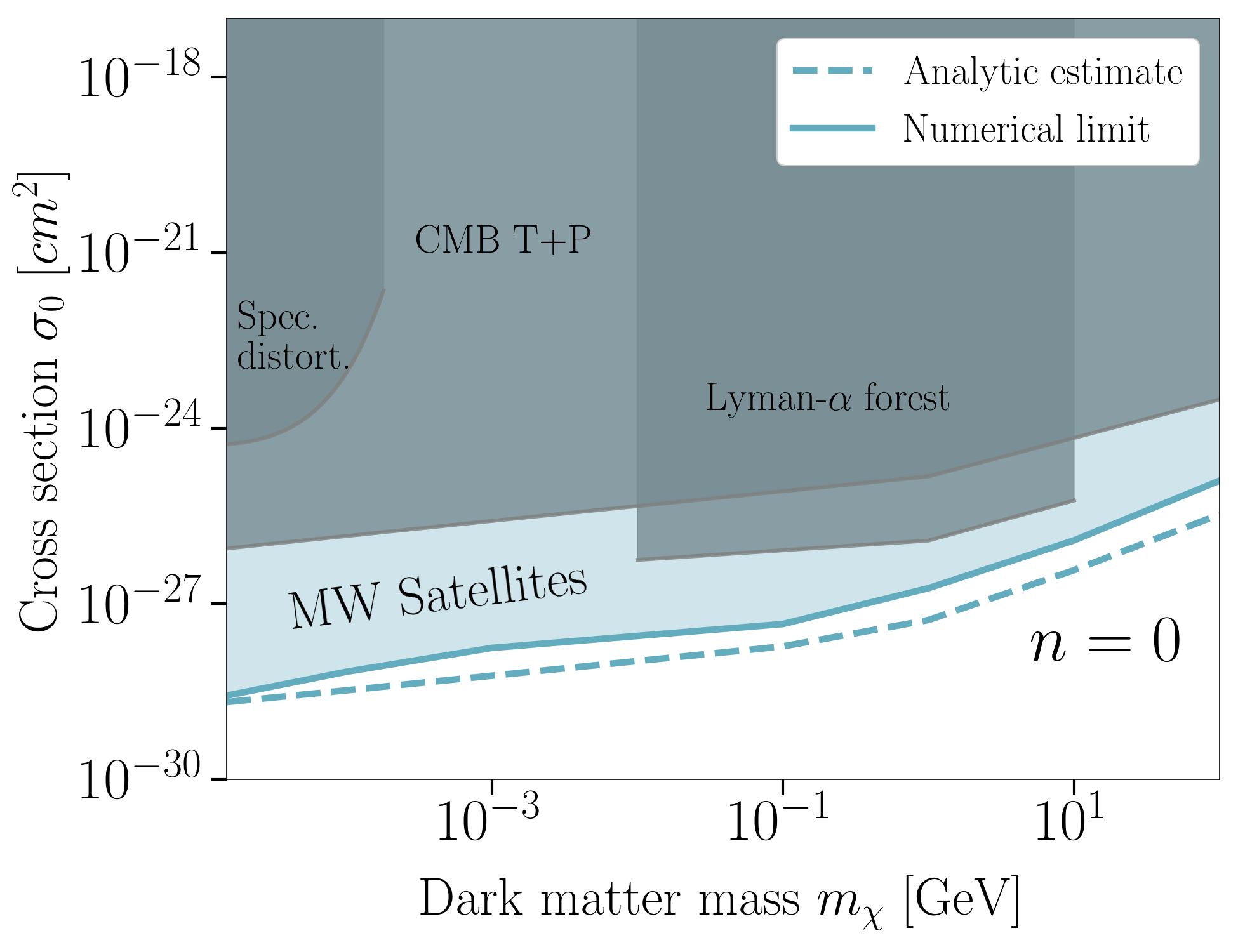}
    \includegraphics[width=0.475\textwidth]{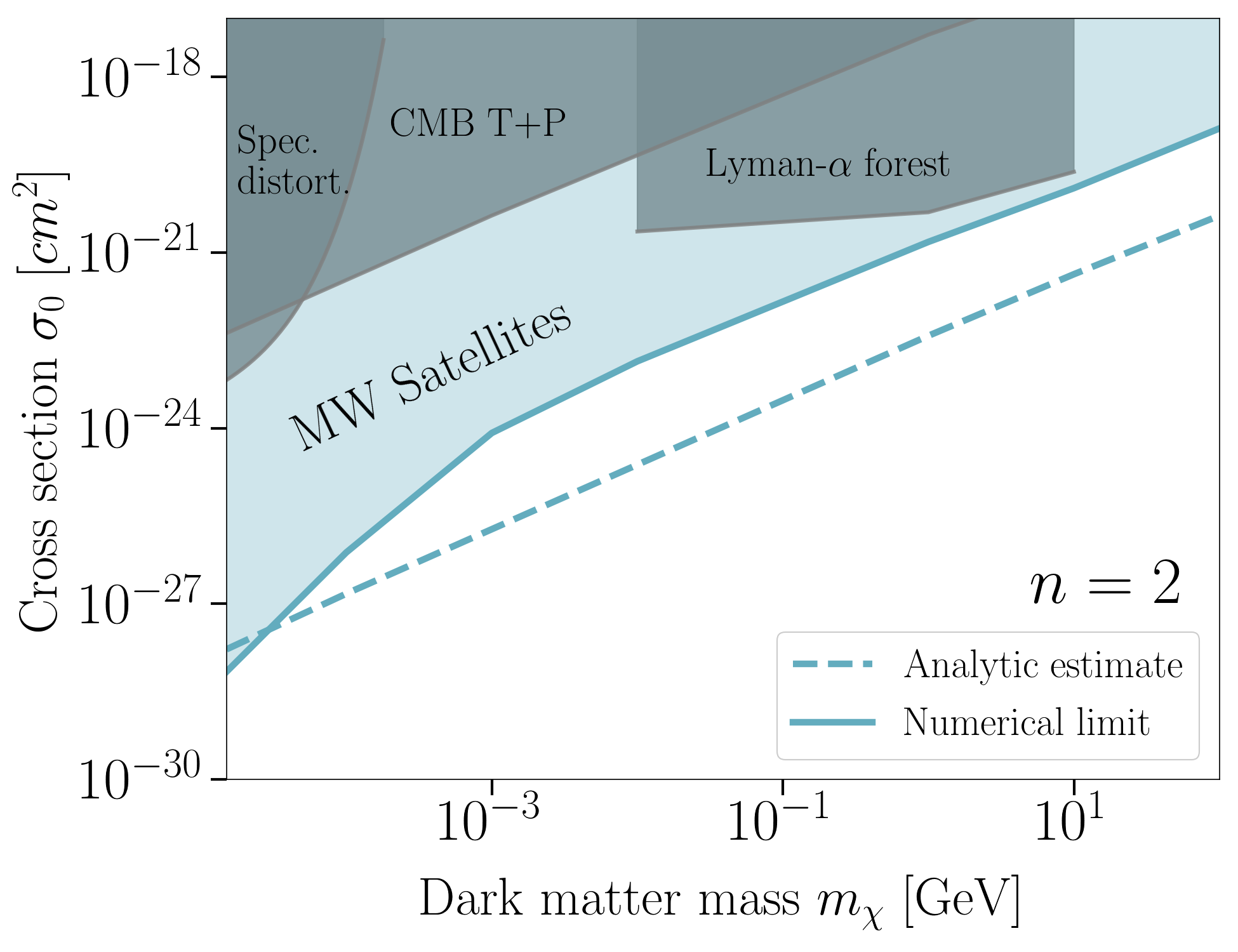}
    \vspace{3mm}
    \includegraphics[width=0.475\textwidth]{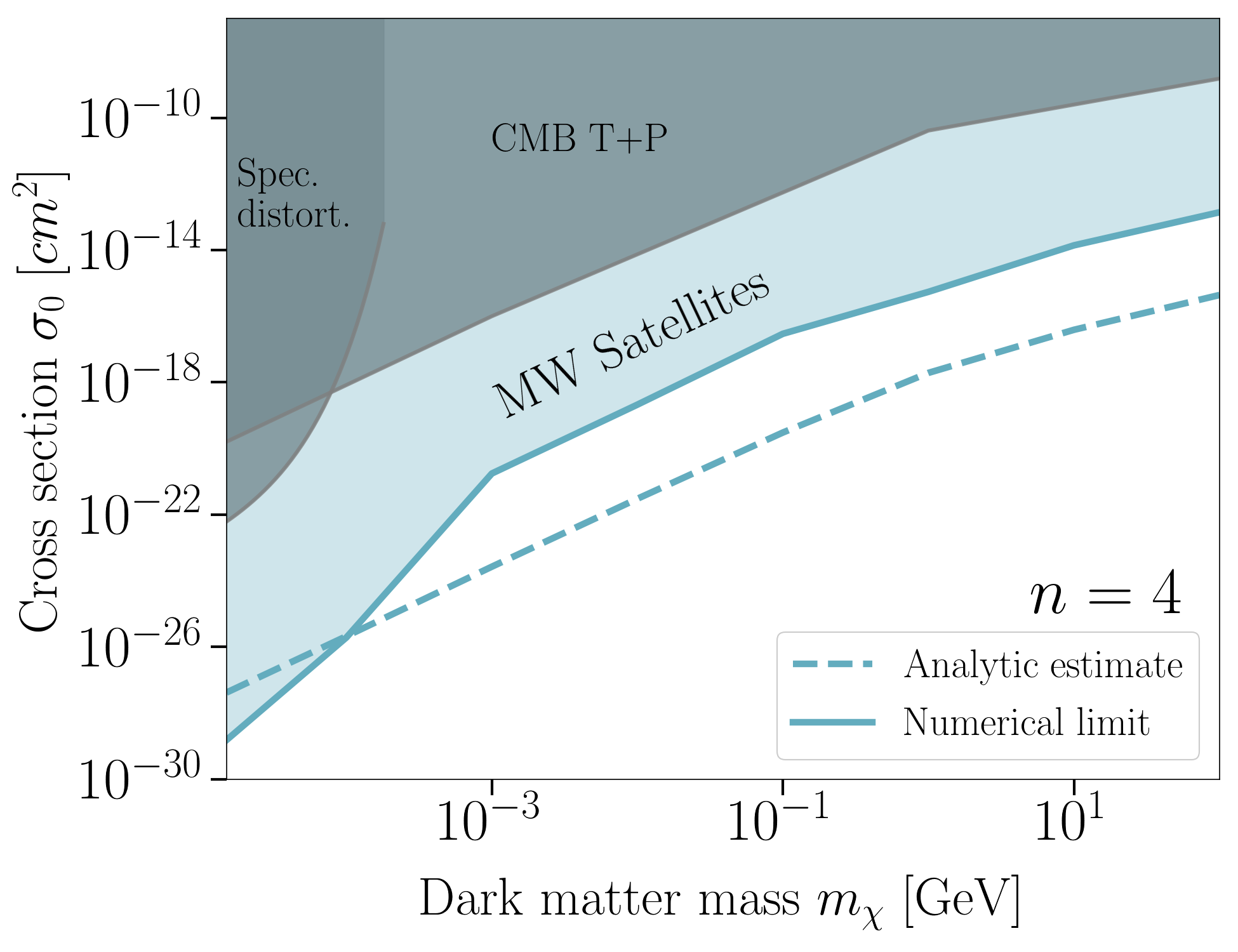}
    \includegraphics[width=0.475\textwidth]{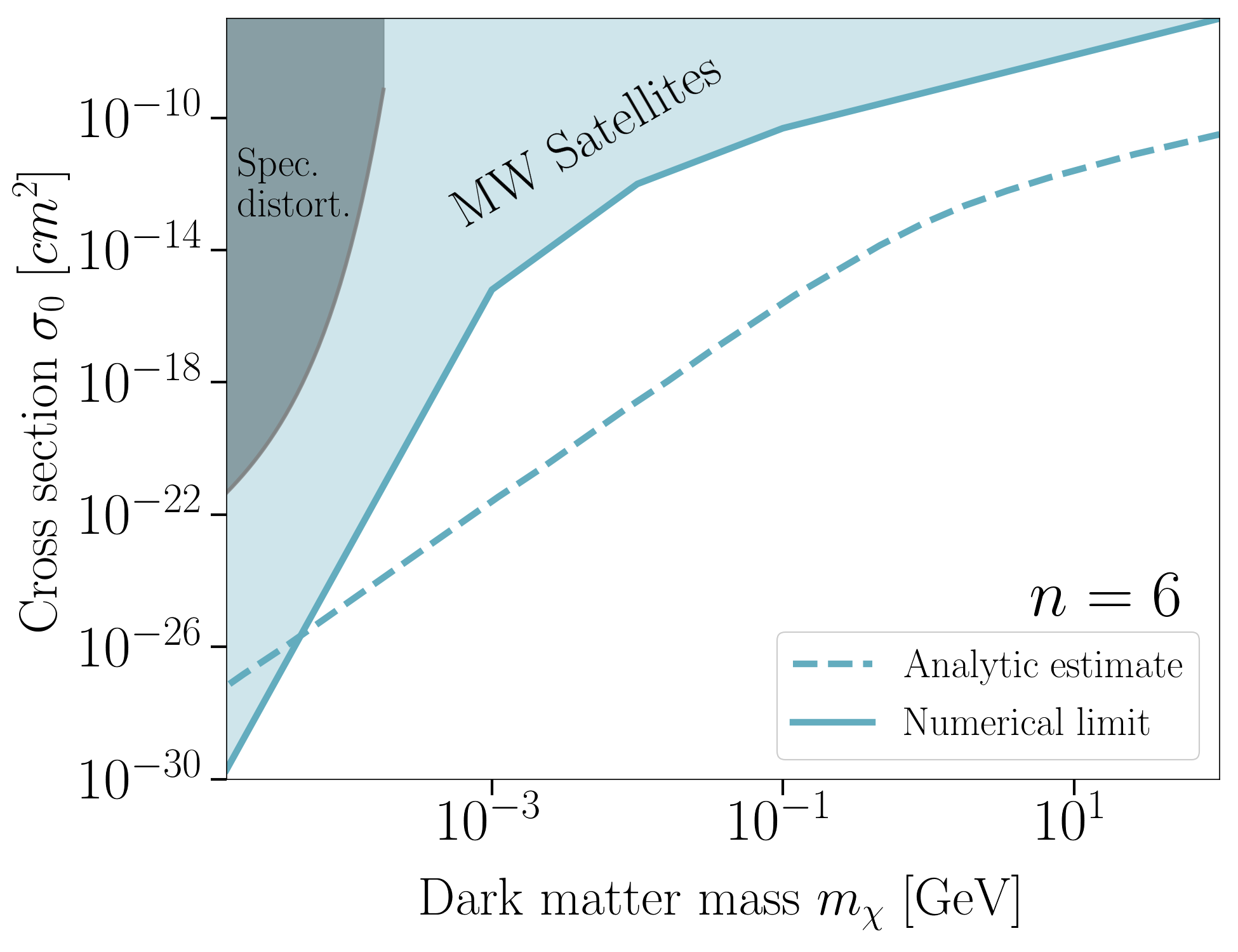}
    \caption{Upper bounds on DM--proton scattering cross section. Each panel corresponds to a different velocity dependence of the DM--baryon interaction model, specified by the power-law index $n$ (shown in the lower right of each panel). Solid blue lines indicate our numerical limits; the blue shaded regions of the parameter space are inconsistent with the observed abundance of Milky Way satellite galaxies from DES and PS1 data \citep{Nadler:2020}. 
    Dashed blue lines indicate our analytic estimates of the same bounds.
    Where available, we show the corresponding limits from the \textit{Planck} temperature and polarization anisotropy measurement \citep{BoddyGluscevic}, from spectral-distortion bounds from FIRAS \citep{AliHaimoud}, and from Lyman-$\alpha$ forest measurements \citep{Xu189710}, as grey shaded regions.
    For each interaction model, we report orders of magnitude of improvement over previous bounds.}
    \vspace*{0.2cm}
    \label{fig:limits}
\end{figure*}
%%%%%%%%%%%%%%%%%%%%%%%%%%%%%%%%%%%
Our numerical bounds on $\sigma_0$ as a function of $m_\chi$ are presented in Figure~\ref{fig:limits} and Table \ref{tab:limits} for $n\in\{0,2,4,6\}$.
In the same Figure, we present the results of our analytic estimates, cast as an equivalent limit on $\sigma_0$.
We also show the previous limits from \textit{Planck} measurements of the CMB temperature and polarization anisotropy \citep{BoddyGluscevic}, the limits from FIRAS spectral-distortion bounds \citep{AliHaimoud}, and the limits from Lyman-$\alpha$ forest analysis \citep{Xu189710}.
Our numerical limits are orders of magnitude more constraining than those in previous studies and currently present the most stringent astrophysical bounds on these IDM models. 
Comparing to the \textit{Planck} limits, we report an improvement of approximately 3 and 5 orders of magnitude for $n=2$ and $n=4$, respectively. 
For $n=0$, our findings are consistent with \cite{NadlerGluscevic,NadlerDES}.
%%%%%%%%%%%%% Table of limits 
\begin{table}[t]
\centering
\begin{tabular}{| c || c | c | c | }
 \hline
 \textbf{n} & \textbf{Mass} &
 \textbf{Numerical limit} &
 \textbf{Analytic estimate} \\
 & & \textbf{[cm$^2$]} & \textbf{[cm$^2$]} \\
 \hline \hline
   & $15$ keV & $2.7\times10^{-29}$ & $2.1\times10^{-29}$ \\
   & $100$ keV & $6.9\times10^{-29}$ & $3.3\times10^{-29}$  \\
 $0$ & $10$ MeV & $2.8\times10^{-28}$ & $1.0\times10^{-28}$\\
   & $1$ GeV & $1.8\times10^{-27}$ & $5.3\times10^{-28}$\\
   & $10$ GeV & $1.2\times10^{-26}$ & $3.7\times10^{-27}$\\
   & $100$ GeV & $1.3\times10^{-25}$ & $3.3\times10^{-26}$\\
 \hline
   & $15$ keV & $6.9\times10^{-29}$ & $1.7\times10^{-28}$ \\
   & $100$ keV & $7.5\times10^{-27}$ & $1.5\times10^{-27}$\\
 $2$ & $10$ MeV & $1.4\times10^{-23}$ & $2.4\times10^{-25}$\\
   & $1$ GeV & $1.5\times10^{-21}$ & $3.8\times10^{-23}$ \\
   & $10$ GeV & $1.3\times10^{-20}$ & $4.3\times10^{-22}$\\
   & $100$ GeV & $1.3\times10^{-19}$ & $4.3\times10^{-21}$\\
 \hline
   & $15$ keV & $1.5\times10^{-29}$ & $4.2\times10^{-28}$ \\
   & $100$ keV & $1.9\times10^{-26}$ & $2.1\times10^{-26}$\\
 $4$ & $10$ MeV & $2.1\times10^{-19}$ & $3.0\times10^{-22}$\\
   & $1$ GeV & $5.5\times10^{-16}$ & $1.9\times10^{-18}$ \\
   & $10$ GeV & $1.4\times10^{-14}$ & $3.9\times10^{-17}$\\
   & $100$ GeV & $1.4\times10^{-13}$ & $4.4\times10^{-16}$\\
 \hline
  & $15$ keV & $1.9\times10^{-30}$ & $6.6\times10^{-28}$ \\
  & $100$ keV & $7.0\times10^{-24}$ & $2.1\times10^{-25}$\\
 $6$ & $10$ MeV & $1.0\times10^{-12}$ & $2.7\times10^{-19}$\\
   & $1$ GeV & $6.2\times10^{-10}$ & $7.8\times10^{-14}$ \\
   & $10$ GeV & $7.8\times10^{-9}$ & $2.5\times10^{-12}$\\
   & $100$ GeV & $1.0\times10^{-7}$ & $3.2\times10^{-11}$\\
    \hline
\end{tabular}
\caption{Bounds on the normalization $\sigma_0$ of the momentum-transfer cross section, $\sigma_\text{MT}=\sigma_0 v^n$, obtained via the analytic and numerical approaches, for a set of DM masses $m_\chi$ and power-law dependencies on particle velocity $v$, with an index $n$. Table entries correspond to the limits shown in Figure~\ref{fig:limits}.}
\label{tab:limits}
\end{table}

Our numerical limits are the most conservative upper bounds on the momentum-transfer cross section from linear perturbation theory, in the sense that larger values of $\sigma_0$ lead to dramatic decrements in power on scales that are measured to be consistent with CDM.
The analytic estimates, on the other hand, roughly identify values of $\sigma_0$ below which current data has a limited constraining power.
However, the analytic prescription is a poor predictor of the bound at low DM masses for $n>0$ models, as the the analytically--estimated cross sections fall into the excluded regions of the parameter space.

We note that the mass dependence of the numerical bound shown in Figure~\ref{fig:limits} differs from the dependence of the analytic estimate.
While the analytic estimate directly inherits its mass dependence from the DM--baryon momentum transfer rate $R_\chi$, the numerical bound is additionally modulated by the requirement that the DAO features fall strictly beneath the WDM transfer function.
The size of DAO features as a function of $m_\chi$ is not straightforwardly modeled, but it affects the mass dependence of the numerical limit.

%%%%%%%%%%%%%%%%%%%%%%%%%%%%%%%%%%%%%%%%%

\section{Conclusions and Discussion}
\label{sec:conclusion}

We use the latest measurements of the Milky Way satellite population from DES and Pan-STARRS1 to infer the most stringent astrophysical bound to date on velocity-dependent interactions between dark matter particles and protons. 
We generalize methods we previously developed for velocity-independent scattering and apply them to any velocity-dependent interaction that dominates over Hubble expansion in the early Universe.
We do \text{not} assume any specific high-energy behavior of dark matter, and thus probe the parameter space for sub-GeV dark matter in a generic way, complementary to laboratory experiments, providing an important guide for identifying viable candidate models. 
Our results exclude interaction cross sections that can be reached with the next-generation cosmic microwave background experiments \citep{2018PhRvD..98l3524L}.
The methods we developed here are applicable to future data from facilities such as the Vera C.~Rubin Observatory \citep{LSST09120201} and from line-intensity tomography surveys \citep{2020PhRvD.101f3526M}.

We identify a few promising directions for follow-up studies.
First, we note that our numerical results use linear perturbation theory and rely only on the most robust features of the matter transfer function that arise as a consequence of dark matter scattering. 
A simulation-based approach to consistently and fully forward-model satellite populations within an interacting dark matter cosmology can improve upon our results.
Furthermore, the behavior of the dark acoustic oscillations we observe in Figure~\ref{fig:cutoff} may be possible to model semi-analytically to understand their effects on dark matter substructure in galaxies like the Milky Way.
Indeed, such approaches will be necessary to move beyond limits and toward a \textit{discovery} of new dark matter physics, should inconsistencies with the cold dark matter paradigm arise in future measurements.
Simulations that include dark matter--baryon scattering could also uncover other potentially observable signatures of the interactions, such as impacts on halo density profiles.
A combined analysis of all available observational probes is perhaps the most robust way to search for new physics of dark matter with upcoming surveys.

Finally, we note that the validity of our results does \textit{not} explicitly require thermal production of dark matter.
However, we do assume that dark matter follows a Maxwell-Boltzmann distribution, achieved by the sufficiently strong coupling with baryons, through any one of the interactions we considered.
Deviations from this assumption may occur because thermal decoupling takes place before the interactions themselves decouple  \citep{Ali-Haimoud:2018dvo}. 
Furthermore, thermally-produced dark matter can alter primordial element abundances \citep{Boehm13036270,Nollett14116005,Krnjaic190800007} and there are corresponding limits on thermal dark matter candidate mass, complementary to our results.
However, these limits rely on details of a specific model of dark matter, such as its spin statistics and the high-energy behavior of its interactions; in contrast, we need not make any assumptions about these details.
We thus leave detailed comparisons of these bounds and considerations related to the velocity distributions for the future.

%%%%%%%%%%%%%%%%%%%%%%%%%
\section{Acknowledgments}

K.~M.~acknowledges the support through the Provost's Research Fellowship for undergraduate students at USC. V.~G.~is supported by the National Science Foundation under Grant No. PHY-2013951. E.~O.~N.~is supported by the National Science Foundation (NSF) under grant No.\ NSF DGE-1656518 through the NSF Graduate Research Fellowship. The authors thank Yacine Ali-Ha\"{i}moud and Keith Bechtol for useful comments.

%%%%%%%%%%%%%%%%%%%%%%%%%%%%%%%%%%%%%%%%%%%%%%%%%%%%
%\clearpage
\bibliographystyle{yahapj}
\bibliography{biblio}

\end{document}